\documentclass[nohyper,12pt,letterpaper]{JHEP3}
\usepackage{epsfig}
\usepackage{amsmath}
\usepackage{subfigure}

\DeclareSymbolFont{AMSa}{U}{msa}{m}{n}
\DeclareSymbolFont{AMSb}{U}{msb}{m}{n}
\let\Box\relax
\DeclareMathSymbol{\Box}{\mathord}{AMSa}{"03}

\newcommand{\be}{\begin{equation}}
\newcommand{\ee}{\end{equation}}
\newcommand{\bea}{\begin{eqnarray}}
\newcommand{\eea}{\end{eqnarray}}

\title{Hierarchies of Susy Splittings and Invisible Photinos as  Dark Matter}
\author{W. Fischler${}^{}$\\
Department of Physics and Texas Cosmology Center\\ The University of Texas at Austin,
TX 78712.
\\ E-mail: \email{fischler@physics.utexas.edu}}

\author{W. Tangarife Garcia${}^{}$\\
Department of Physics and Texas Cosmology Center\\ The University of Texas at Austin,
TX 78712.
\\ E-mail: \email{wtang@physics.utexas.edu}}

\abstract{ We explore how to generate hierarchies in the splittings between superpartners. Some of the consequences are the existence of invisible components of dark matter, new inflaton candidates, invisible monopoles and a number of invisible particles that might dominate during various eras, in particular between BBN and recombination and decay subsequently. }


\received{???????? ?st, 2010} \accepted{???????? ?th, 1998}
\preprint{
UTTG-13-10\\TCC-027-10}

\begin{document}




\section{\bf Introduction}


In this paper we will explore how to generate hierarchies in the splittings between super-partners.   We will be mostly  focusing on the cases where the hierarchy is such that it generates a sector with almost degenerate super-partners.  Producing such an hierarchy will be done in the context of gauge mediation. The ``almost supersymmetric" sector  will be largely decoupled from the visible degrees of freedom and this decoupling will generically increase as the invisible sector becomes more and more supersymmetric. 

Our discussion will remain in a field theoretical context, we will not discuss here the stringy realization of such a hierarchy.

As expected, the invisible sector that emerges in this construction does generate possible candidates for dark matter.
Present observations have not yet revealed the nature of the dark matter which remains a mystery. Various proposals for the nature of dark matter exist. These proposals run the gamut from ``weakly" interacting particles to ``invisible" degrees of freedom (see  \cite{dm} to \cite{feng}). A substantial number of experiments are devoted to finding what makes up the dark matter (\cite{ab} to \cite{ab-end}). The kind of dark matter considered in this paper is of the ``invisible" kind. We will consider ranges of parameters where the partner of an invisible photon is the main component of dark matter. In a different range of parameters invisible monopoles magnetically charged for the invisible photon will dominate.

Another possible effect the invisible sector can produce is the existence of a light non relativistic invisible particle that dominates the energy density of the universe between BBN and recombination and then decays \footnote{In a separate paper \cite{ad},  observational bounds for such a matter dominated era from the fluctuation spectrum of the microwave background are discussed.}. This is an era where radiation domination is assumed and where there is little in ways to check what the actual equation of state is \cite{linder}.

In addition, features of the construction presented here include the existence of inflaton candidates with extremely flat potentials. As it turns out, these inflatons still do require fine tuning to generate the right amplitude for the fluctuations observed in the microwave background.

The paper is organized as follows: the first section will describe the general construction, in the context of gauge mediation, of hierarchies in the splittings between super-partners. In section 2, for illustrative purposes we focus on a specific model realizing such a hierarchy and we discuss the associated spectra of particles. In section 3, we present various regions of parameters and the cosmological implications of the invisible sector. We close with conclusions.

\section{\label{hierarchies}\bf Hierarchies of Susy Splittings}

\begin{figure}[h]
\begin{center}
\includegraphics[width=5cm]{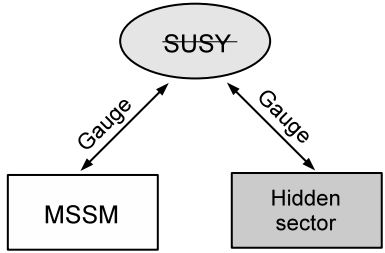}
\end{center}
\caption{Scheme of a model with gauge mediation and a hidden sector.}
\label{model0}
\end{figure}

For the sake of illustrating the general idea, we will mostly focus in this paper an O'Raifeartaigh-like superpotential \cite{OR} for which SUSY is spontaneously broken at some scale $\mu$ and its breaking is mediated via gauge interactions (GMSB) (see \cite{gmsb} to \cite{gmsb-end}). This can easily be generalized. Specifically, the model will have what we will dub a visible gauge group and an invisible gauge group. The model has two flat directions $X$, $Y$, and four messenger multiplets $A$, $B$, $C$ and $D$. $A$ and $B$ are chiral multiplets in the adjoint representation of the visible sector gauge group. $C$ and $D$ are chiral multiplets in the adjoint representations of the visible sector gauge group and the invisible gauge group of some hidden sector that does not couple at tree level with the visible one. The fact that $C$ and $D$ are charged under both gauge groups will be important for the reheating process discussed in section \ref{reh}. $X$ and $Y$ are singlets under both gauge groups (see figure \ref{model0}). The superpotential for the messenger sector is given by 
\begin{eqnarray}
W_{\rm mess}\,&=&\,Y\,{\rm Tr}[\lambda_1 A^2\,+\,\lambda_2 C^2\,-\,\mu^2]\,+\,X\,{\rm Tr}[\kappa_x C^2]\,+\,m_1\,{\rm Tr}[A\,B]\nonumber \\&&+m_1\,{\rm Tr}[A^2]\,+\,m_2\,{\rm Tr}[C\,D]\,+\,m_2\,{\rm Tr}[C^2]. \label{messenger}
\end{eqnarray}
For this superpotential, the $F$ equations are given by
\begin{eqnarray}
F^{*}_Y&=&{\rm Tr}[\lambda_1 A^2+\lambda_2 C^2 -\mu^2], \nonumber \\
F^{*}_X&=&{\rm Tr}[\kappa_x C^2], \nonumber \\
F^{*}_A &=& 2\lambda_1 Y A+m_1 B +2m_1A, \nonumber\\
F^{*}_B&=& m_1A, \nonumber \\
F^{*}_C&=& 2 \lambda_2 Y C+2 \kappa_x X C+m_2D+2m_2C, \nonumber\\
F^{*}_D&=&m_2 C. \label{F-eq}
\end{eqnarray}
The messenger contribution to the $D$-term equations is given by 
\be
D^a=\frac{g^2}{2}{\rm Tr}(T^a[\phi^\dagger,\phi]),\,\,\,\,\,\,\phi\,=\,A,\,B,\,C,\,D;
\ee where $T^a$ is the generator of the respective gauge group and $g$ is the gauge coupling constant.

We notice that for such a set of equations, there is no solution for $F^{*}_i=0$ $\forall\,\,i$. Therefore there is no supersymmetric vacuum. Meaning the absence of $V_{min}=0$ for the scalar potential
\be
V=\sum_i|F_i|^2\,+\,\frac{1}{2g^2}\sum_a|D^a|^2. \label{V}
\ee
Instead, for vanishing expectation values of the messenger fields, the global minimum of the potential is $V_{min}=\mu^4$ \footnote{This approach may be extended to generalized gauge mediation (\cite{gmsb},\cite{sg} to \cite{sg-end}). Other discussions about supersymmetry breaking in hidden sectors can also be found in \cite{mc}.}.  This supersymmetry breaking implies a splitting in the superpartner masses for the spectrum in the visible and invisible sector. Taking $m_1\,<\,m_2$, implies a hierarchy between the mass splittings 
\be
\delta m^2_{\rm visible}\,\approx\,\frac{\alpha^2\lambda_1^2}{4\pi^2}\frac{\mu^4}{m_1^2}\,\,\gg\,\, \delta m^2_{\rm invisible}\,\approx\,\frac{\alpha_h^2\lambda_2^2}{4\pi^2}\frac{\mu^4}{m_2^2}.\label{mass-splitting}
\ee

Notice that $\mu$ and $m_1$ are constrained by the requirement of $\delta m_{\rm visible}$ being of the order of the electroweak scale. Therefore,

\be
\mu^2\,\approx\, \frac{4\pi^2\,m_1\,(10^3\,\,{\rm GeV})}{\lambda_1\,\alpha}. \label{ew-condition}
\ee

The mass splitting hierarchy discussed above is easily generalized to models in which there are $n$ ``hidden" sectors that interact through gauge interactions with the messenger sector, by endowing a set of messengers with a mass hierarchy 
\be
m_1\,<m_2\,<... <m_n.
\ee

\section{\bf More about this model}
\subsection{\bf Inflation}
\begin{figure}[h]
\begin{centering}
\begin{center}
\includegraphics[width=6cm]{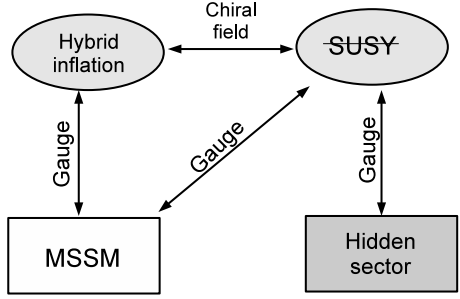}
\end{center}
\end{centering}
\caption{Extended model scheme.}
\label{model}
\end{figure}

The model we discussed above could, in principle, use the $X$ field as a candidate for inflation since $m_2 \,>\,m_1$ implies that the effective potential due to 1-loop quantum corrections \cite{C-W}, captured in equation (\ref{CW}) below, is very flat. However, when we compute the fluctuations in the energy density due to the inflaton, their magnitude is actually very small, $$\frac{\delta \rho}{\rho}\,\sim\,10^{-14},$$ as we keep the coupling constant $\lambda_1\approx 1$. To be consistent with the experimental value \cite{wmap}, we would have to fine tune this coupling to be really small. 
In order to include a model for inflation devoid of this challenge, we will extend the superpotential in equation (\ref{messenger}) and add a $W_{\rm inflation}$: 
\be
W_{\rm inflation} \,=\,S\,{\rm Tr}[\kappa\, \Phi^2\,-\,\mu_1^2]\,+\,{\rm Tr}[\xi\,\Phi\,C^2],\label{hybrid}
\ee
where $\Phi$ is a chiral superfield in the adjoint representation of the visible and invisible gauge groups and S is a singlet. In figure (\ref{model})  we present the schematics of this extended model. This kind of inflationary model is known as hybrid inflation \cite{linde, dvali}.  We notice that the scalar potential (\ref{V}) for this model has a supersymmetric minimum at $ \langle S \rangle = 0$, $\langle C \rangle =0$ and $ \langle \Phi \rangle = \pm \mu_1^2/\kappa$. However, for $S > S_c=\mu_1/\sqrt{\kappa}$, the $\Phi$ field stays at the origin and SUSY is spontaneously broken with $S$ as a flat direction. This vacuum degeneracy is lifted by quantum corrections, for which we obtain the Coleman-Weinberg potential \cite{C-W} \cite{intri-seiberg}
\begin{equation}
V^{(1)}\,=\,\frac{1}{64\pi}\left( {\rm Tr}\left[m_B^4{\rm ln} \frac{m_B^2}{\Lambda^2}\right]-{\rm Tr}\left[m_F^4{\rm ln} \frac{m_F^2}{\Lambda^2}\right]\right),\label{CW}
\end{equation}
where  $m_B$ and $m_F$ are the mass matrices for bosonic and fermionic fields respectively.

Thus, we can write the effective potential (\ref{CW}) for hybrid inflation when $S \gg S_c $:
\begin{equation}
V^{(1)}(S)\,\approx\,\mu_1^4 \left[1+\frac{\kappa^2}{8\pi^2}{\rm ln}\left(\frac{S}{S_c}\right).\right]
\end{equation}

For large values of $S$, this potential satisfies the slow roll conditions. The inflationary stage ends as $S\approx S_c$ after which value the $\Phi$ fields roll towards the supersymmetric vacuum.  The number of e-foldings during inflation is given by \cite{weinbergcosmo} 
\begin{equation}
\mathcal{N}\,=\,-\int_{S_N}^{m_2/\kappa}dX\, \frac{8\pi}{M_p^2}\frac{V(X)}{V'(X)}\approx \frac{64\pi^3}{\kappa^2\,M_p^2}[S_N^2-S_c^2] \label{N}
\end{equation} and the density perturbations are 
\begin{equation}
\frac{\delta\rho}{\rho}\,=\,4\sqrt{\frac{8\pi}{75}}\frac{V^{3/2}}{M_p^3\,V'}\,\approx\,\sqrt{\frac{8\pi}{75}}\frac{32\pi\,\mu_1^2S_N}{\kappa^2\,M_p^3}.\label{pertu2}
\end{equation}

With the choice $\kappa\approx6.0 \times 10^{-2}$, SUGRA effects can be neglected \cite{linde-riotto}, and using the energy density perturbations  $\delta\rho/\rho\approx 10^{-5}$ \cite{wmap}, we see that $\mu_1\,\approx\,3.2\times10^{15}$ GeV  if one rquires $N\sim 60$. For this set of parameters, we obtain a ``red" spectral index $n_s\,\approx$  0.98. Other results for larger values of $\kappa$ have been discussed in \cite{linde-riotto}. 

\subsection{\bf Light degrees of freedom in the hidden sector}
Now, let us consider the ``invisible" sector. Again for simplicity, the degrees of freedom will consist of a chiral multiplet $H$ in the adjoint representation of  the gauge group $G_{hidden}$, chosen to be $SU(2)_h$. The spectrum also includes a $SU(2)_h$ singlet field $Z$. The superpotential for these fields is 
\be
W_{\rm hidden}\,=\,Z\,{\rm Tr}[\epsilon H^2\,-\,v_h^2],\label{hidden}
\ee where $\epsilon$ is a coupling constant and $v_h$ is some energy scale. Throughout this  paper,  we assume canonical K\"ahler potentials for all the fields. At the minimum of the invisible sector potential, $SU(2)_h$ is broken to $U(1)_h$ by the expectation values
\be
\langle \overrightarrow{H} \rangle\,=\,(0,\,0,\,v_h/\sqrt{\epsilon}), \,\,\,\,\,\,\,\,\langle Z \rangle\,=\,0. 
\ee
The resultant mass spectrum is
\be m_{H_3}^2\,=\,4\epsilon\,v_h^2,\,\,\,m^2_Z\,=\,4\epsilon\,v_h^2,\,\,\,m^2_{W_\pm}\,=\,g_h^2v_h^2,\,\,\,\,m^2_{W_3}\,=\,0.
\ee 
In addition, the scalar field $\tilde{H_3}$  receives a two-loop mass contribution due to supersymmetry breaking in the messenger sector as mentioned in equation (\ref{mass-splitting}). The gaugino fields receive a one-loop mass contribution while the $\tilde{Z}$ mass splitting comes from three-loop diagrams. As we saw in section \ref{hierarchies}, these mass splittings are much smaller than the splittings in the visible sector. 
\be
\delta m^2_{ \tilde{H_3}}\,\approx\,\frac{\alpha_h^2\lambda_2^2C_2}{4\pi^2}\frac{\mu^4}{m_2^2},\,\,\,\,\delta m_{\tilde{W_i} }\,\approx\,\frac{\alpha_h\lambda_2 C_2}{8\pi}\frac{\mu^2}{m_2},\,\,\,\,\delta m^2_{\tilde{Z}}=\frac{\epsilon^2\alpha_h^2\lambda_2^2C_2}{16\pi^4}\frac{\mu^4}{m_2^2}. \label{splitting-hidden}
\ee

\begin{figure}[h]
\begin{centering}
\begin{center} 
\includegraphics[width=9cm]{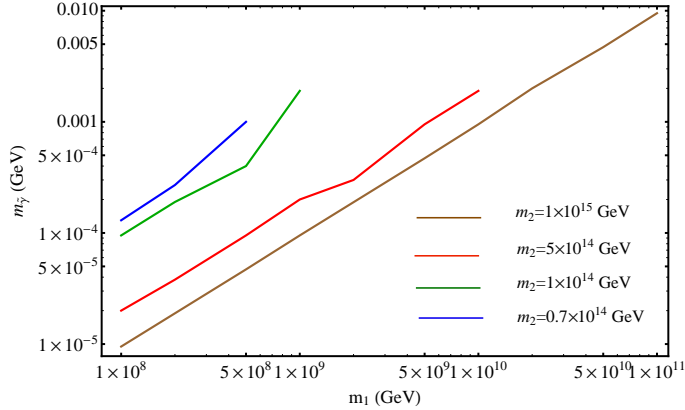}
\end{center}
\end{centering}
\caption{Mass of the photino for different values of $m_1$ and $m_2$ in the ranges determined in section \ref{relic-abund}.}
\label{mph}
\end{figure}

The $Z$ and $H$ particles of this sector rapidly decay  into $W$ fields or radiation. On the other hand, the lightest particle in this sector is the ``invisible" photino $\tilde{W_3}$ (or $\tilde{\gamma}$), whose mass comes solely from the supersymmetry breaking mediation. As the hidden sector is ``almost" supersymmetric, this photino turns out to have a  small mass $10^{-5}\,\, {\rm GeV} \le m_{\tilde{\gamma}} \le 10^{-2} \,\,{\rm GeV}$ whose range depends on the values of $m_1$ and $m_2$ as shown in  figure (\ref{mph}). This particle can decay only into an ``invisible" photon and a gravitino.  
 
 There are no observable signals in the visible world of the decays or annihilations of invisible particles. This is due to the fact that the two sectors are largely decoupled. As an example, consider  two invisible photons going into two visible photons. This process arises from a dimension 8 operator 
 \be 
  F_h^{\mu\nu}F_{h,\mu\nu}F^{\rho\sigma}F_{\rho\sigma}, \label{mix}
 \ee 
 which is suppressed by a quite small coefficient $\sim m_2^{-4}$. Also, the gauge group for the hidden sector is asymptotically free. On the other hand, the new content of fields makes the visible world beta functions positive, such that these interactions are not asymptotically free for scales larger than $m_1$ although they stay in the perturbative regime for energies below the Planck scale.
 
 However, the invisible sector has  cosmological implications which will be discussed in the next section. 
 
\section{\bf Cosmological implications of the invisible sector}
\subsection{\label{reh} \bf Reheating process}
After inflation ends, as will be seen in what follows, the visible and invisible sectors will reheat. When the inflaton $S$ rolls down to the supersymmetric vacuum, the $\Phi$ fields oscillate and decay into visible and invisible radiation. This decay happens quite early and radiation dominates the universe. When $H=\dot{a}/a$ is small enough, $X$ and $Y$ (that were frozen during inflation at some sub-Planckian value far from the origin\footnote{ Because of inflation, these fields can be frozen at  any value in field space, it is not improbable that they are frozen at values far from the origin.}) roll down towards the origin and start oscillating. At some time $t_0$ their oscillations dominate the energy of the universe and there is a transition to a matter dominated expansion of the universe. The $Y$ field decays into the visible sector faster than the $X$ field, because of  $ m_1< m_2$, which is responsible for the subsequent hierarchy among superpartners..
The decay rates are

\be
\Gamma_Y \sim \frac{\alpha^2C_2^2\,m_Y^3}{m_1^2},\,\,\,\,\Gamma_X \sim \frac{\alpha^2C_2^2\,m_X^3}{m_2^2},
\ee
where $C_2$ is the Casimir of the gauge group. The $X$ field also decays into the hidden sector with a decay rate similar to its decay in the visible sector. This decay, $\Gamma_X^h$,  one has to replace in this channel, $\alpha$ by the coupling constant of the {\it invisible} gauge group $\alpha_h$. In equation (4.1), $m_Y$ comes from the 1-loop $Y$ mass contribution 
\be
m_Y^2\,\approx\,\frac{\lambda_1^4}{16\pi^2}\frac{\mu^4}{m_1^2}, 
\ee which turns out to be dependent on $\lambda_1$ and $\alpha$ according to equation (\ref{ew-condition}). On the other hand, $m_X$ comes from a tree-level term $m_X X^2$ that is added to the superpotential in equation (\ref{messenger}) so that $X$ is not very long-lived.   

After $t_0$, the radiation energy densities $\rho_R$ and $\rho_{R,h}$ in both sectors are given by the solutions to the equations \cite{kolb-turner}
\begin{eqnarray}
\dot{\rho}_R+4\,H\,\rho_R&=&\Gamma_Y\,\rho_Y+\Gamma_X\,\rho_X \label{rho}\\
\dot{\rho}_{R,h}+4\,H\,\rho_{R,h}&=&\Gamma^I_X\,\rho_X. \label{rhoi}
\end{eqnarray}

The solutions for these equations, for $t\le\tau_X$, are 
\bea
\rho_R(t)\,&=&\,\beta\,\rho_{R,\,0}\,\left[\frac{t_0}{t}\right]^{8/3}\,+\,\frac{5}{3}\beta\,\rho_{X,\,0}\left(\frac{t_0^2}{t\,\tau_X }\right)\,+\,\frac{5}{3}\rho_{Y,\,0}\left(\frac{t_0^2}{t\,\tau_Y }\right) \\
\rho_{R,\,h}(t)\,&=&\,(1-\beta)\,\rho_{R,\,0}\,\left[\frac{t_0}{t}\right]^{8/3}\,+\,\frac{5}{3}(1-\beta)\,\rho_{X,\,0}\left(\frac{t_0^2}{t\,\tau_{X,\,h} }\right), 
\eea where $$\beta\,=\,\frac{\Gamma_X}{\Gamma_X^h+\Gamma_X}.$$
After $t=\tau_X$, $\rho_X$ drops exponentially and the ``invisible" radiation density is given by 

\be
\tilde{\rho}_{R,\,h}(t)\,=\,\rho_{X,0}\,(1-\beta)\,\left(\frac{5}{3}+e^{-\frac{t}{\tau_X}}\right)\frac{t_0^2\tau_X}{\tau_{X,\,h}t^2}.
\ee

\begin{figure}[h]
\begin{centering}
\begin{center}
\subfigure[ ]{
\includegraphics[width=7.4cm]{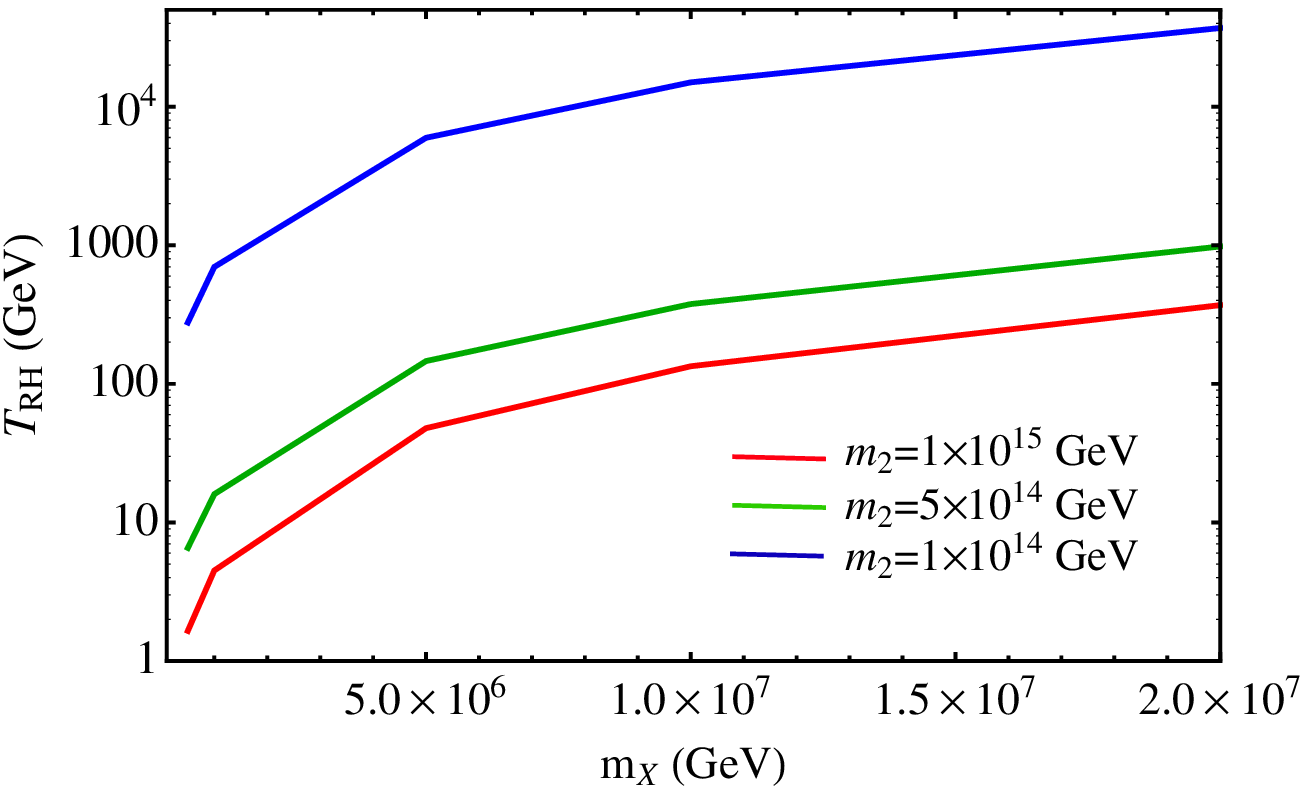}}
\subfigure[ ]{
\includegraphics[width=7.0cm]{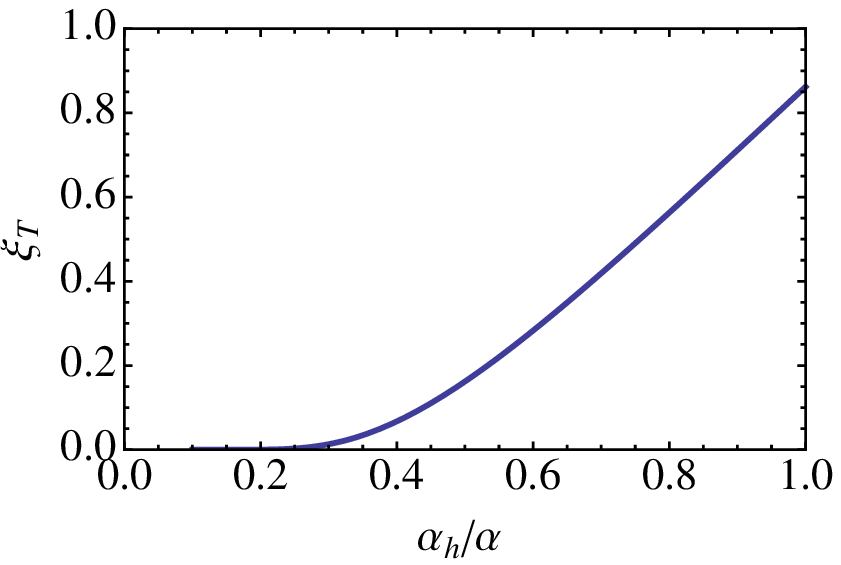}}
\end{center}
\end{centering}
\caption{(a) Range of values for $T_{RH}$. (b) The ratio of temperatures depends on the ratio of the coupling strength between the messenger sector and the visible or hidden sector.} \label{trh}
\end{figure}

The corresponding reheating temperature for each of the sectors is given by 
\be 
T_{RH}\,=\,g_*^{-1/4}(\rho_R)^{1/4},\,\,\,\,\,T_{RH,_h}\,=\,g_{*,h}^{-1/4}(\rho_{R,\,h})^{1/4}.
\ee  In figure (\ref{trh}a) we present the values that $T_{RH}$ can take depending on $m_X$ and $m_2$. For low values of $T_{RH}$, the Affleck-Dine mechanism might be used for baryogenesis \cite{aff-dine}.   

We define the ratio of temperatures as $\xi_T \equiv T_{RH,h}/T_{RH}$. This ratio depends mainly on the ratio of strength of the gauge couplings (evaluated at the $X$ field mass)  between the $C$ field and both sectors $(\alpha_h/\alpha)$. This dependence is represented in figure (\ref{trh}b).  As BBN imposes the  constraint \cite{bbn}
\be
g_{*,h}(T_{BBN,h})\left(\frac{T_{BBN,h}}{T_{BBN}}\right)^4\,\le\,2.45 (95\% {\rm CL}),
\ee we may take $\xi \le 0.9$, since at $T_{BBN}\sim 1$ MeV the only light degrees of freedom in the hidden sector come from the photon and, maybe, the photino. For simplicity, through the remainder of this paper, we will assume $\xi_T \approx 0.3$. 

\subsection{\label{relic-abund}\bf Relic abundance}
For some range of parameters, see below, the photinos in the invisible sector are stable particles. As the universe expands, they ``freeze out" when $n\,\langle \sigma v \rangle \approx H$, where $n$ is the photino number density and $\langle \sigma v \rangle$ is the thermally averaged cross-section. This happens at some temperature $T_{f,h} $ in the invisible sector (or $T_f\,=\,T_{f,h}/\xi_T$ in the visible sector). The thermal relic density is given by \cite{kolb-turner} \cite{feng}  

\be
\Omega_{\tilde{\gamma}}\,\sim\,\frac{x_f\,T_0^3}{\rho_c\,M_P\,\langle \sigma v \rangle}\label{relic}
\ee
where $\rho_c$ is de critical density and $T_0$ is the current temperature of the universe. $x_f\equiv m_{\tilde{\gamma}}/T_{f,h}$ is typically $\sim 20$. 

Given the above considerations, we can use the fact that \cite{wmap}
\be
\Omega_{DM}\,\approx\,0.227 \label{DM}
\ee
to constrain the parameters of our model.

\begin{figure}[h]
\begin{centering}
\begin{center}
\subfigure[ ]{
\includegraphics[width=7.2cm]{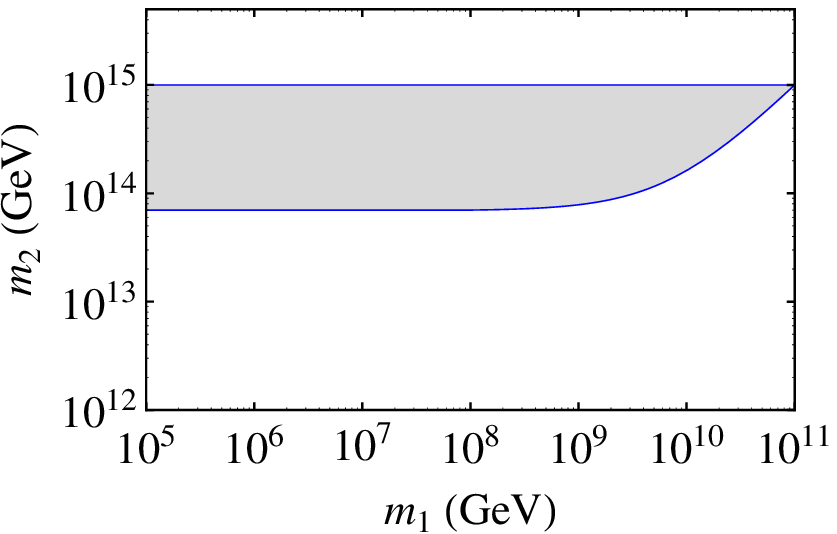}}
\subfigure[ ]{
\includegraphics[width=7.2cm]{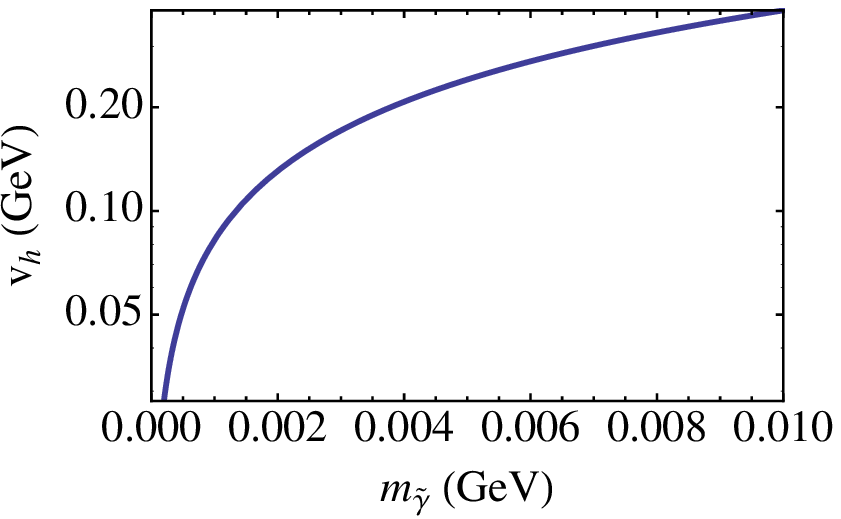}}
\end{center}
\end{centering}
\caption{(a) Range of allowed values of $m_1$ and $m_2$ imposed by the invisible photino decay. (b) Value of the gauge symmetry breaking scale for a given value of the DM photino mass.} \label{m1m2}
\end{figure}

If the invisible photinos are assumed to be candidates for dark matter, their lifetime (before decaying into photon and gravitino) should be longer than the age of the universe. This imposes the constraints  $m_1 \le 10^{11}$ GeV and $7\times 10^{13}\,{\rm GeV}\le  m_2\le 10^{15} \,{\rm GeV}$ as shown in figure (\ref{m1m2}a). Also, we need to constrain the values of the gauge symmetry breaking scale $v_h$ since the photino annihilation cross-section is given by 
\be
\langle\sigma v \rangle \sim \frac{\alpha_h m^4_{\tilde{\gamma}}}{4\pi^3v_h^6}.
\ee Therefore, for photino DM with a given $m_{\tilde{\gamma}}$, we have a specific value of $v_h$ as we show in figure (\ref{m1m2}b). For  $ 10^{-5} \,\,{\rm GeV} < m_{\tilde{\gamma}} < 10^{-2} \,\,{\rm GeV} $ we obtain that $10^{-3} \,\,{\rm GeV} < v_h < 3 \times 10^{-2} \,\,{\rm GeV}$.

Notice that, since $m_{\tilde{\gamma}}\neq m_{\rm weak}$ and $g_h\neq g_w$, photinos are not WIMPs; they are closer to the ``WIMPless DM" that was described in \cite{feng-hidden1, feng-hidden2}, although the values for the photino mass are lower than the ones that are considered in that paper. For other classification of DM candidates in hidden sectors coupled to the visible particles, see for example \cite{cheung}. 

It is worth mentioning that in the scenario we described above, the massive $W$'s and the winos will also freeze out, but their abundance would be much smaller than the photino abundance since they have much larger interaction rates. In addition, as $SU(2)$ is broken to $U(1)$, some monopoles are produced, as described by the Kibble mechanism, with a mass $m_{mon}=4\pi v_h/g_h$ \cite{kibble}\cite{kolb-turner}. However, for these monopoles to be suitable candidates for DM the symmetry breaking scale $v_h$ would have to be of the order of $10^{11}$ GeV, which is much higher than the possible reheating temperatures $T_{RH,h}$ that we can get in the invisible sector. 

However in another region of parameters, invisible monopoles can be the dark matter. Indeed, consider the region of parameter space where:
\be 
 m_1 \approx 10^{6} \,{\rm GeV},\,\,\,\,m_2 \le 10^{6}\, {\rm GeV},\,\,\,\,m_X \ge 10^{6}\, {\rm GeV},   
 \ee we can get $T_{RH,h}$ high enough to produce monopoles whose relic abundance could be of the order of the abundance of dark matter.

Another interesting observation of this model is that it can provide an unusual dominant contribution to the energy density between the end of BBN and the start of recombination.  Indeed, the cosmological era between the end of BBN and the start of recombination is usually assumed to  be radiation dominated. Recently, a paper \cite {linder} has shown that this period could not be dominated by an accelerating equation of state $( w< -1/3)$. There is at present no bounds excluding for example a decelerating matter dominated era. The model considered here in the following range of parameters
\be
 m_1 \approx 1.0\times 10^{7} \,\,{\rm GeV},\,\,\, 10^{11} \,\,{\rm GeV}\,<\,m_2\,<5.0\times 10^{11} \,\,{\rm GeV},\,\,\,
\ee  will indeed have the non relativistic invisible photino start dominating the energy density at a temperature about $1\,-\,10$ keV. Subsequently, at $T \approx 1 \,-\,10$ eV, the invisible photino will  decay into an invisible photon and a gravitino.

\section{\bf Conclusions}

In this paper, we showed through examples how to generate a hierarchy of splittings between superpartners. The multiplets that have such hierarchical splittings with the visible sector are indeed invisible in ``low" energy laboratory  experiments at the LHC and others for a wide range of parameters. 

The invisible sector has, as we showed, several implications for cosmology, specifically providing novel candidates for dark matter. We also did describe an instance where an invisible particle the invisible photino
modifies the equation of state dominating the energy density between BBN and the recombination without affecting the subsequent evolution of the universe.
 It would be quite  interesting to see if the ideas presented here can be realized in string theory.

\section{Acknowledgments}

The research of W.F. and W. T. G. was supported in part by the National Science Foundation under
Grant Numbers PHY-0969020 and PHY-0455649. W. F. would like to thank Raphael Flauger for helpful conversations, W. T. G. would like to thank Joel Meyers for the same.

%


\end{document}